\documentclass[a4paper,fleqn,usenatbib]{mnras}

\usepackage{newtxtext,newtxmath}

\usepackage[T1]{fontenc}
\usepackage{ae,aecompl}

\usepackage{graphicx}	
\usepackage{amsmath}	
\usepackage{amssymb}	

\title[Survival of a planet under photo-evaporation]
{Survival of a planet in short-period Neptunian desert under effect of 
photo-evaporation}

\author[D. E. Ionov et al.]{
Dmitry E. Ionov\thanks{E-mail: pereversi@gmail.com},
Yaroslav N. Pavlyuchenkov,
Valery I. Shematovich
\\
Institute of Astronomy, the Russian Academy of Sciences, Moscow,
Russia
}

\date{Accepted XXX. Received YYY; in original form ZZZ}

\pubyear{2018}

\begin{document}
\label{firstpage}
\pagerange{\pageref{firstpage}--\pageref{lastpage}}
\maketitle

\begin{abstract}
Despite the identification of a great number of Jupiter-like and Earth-like 
planets at close-in orbits, the number of ``hot Neptunes'' --- the planets with 
0.6--18 times of Neptune mass and orbital periods less than 3 days --- turned 
out to be very small. The corresponding region in the mass-period distribution was 
assigned as the ``short-period Neptunian desert''. The common explanation
of this fact is that the gaseous planet
with few Neptune masses would not survive
in the vicinity of host star due to intensive atmosphere outflow
induced by heating from stellar radiation. To check this hypothesis we performed
numerical simulations of atmosphere dynamics for a hot Neptune.
We adopt the previously developed self-consistent 1D model of hydrogen-helium
atmosphere with suprathermal electrons accounted. The mass-loss rates as a
function of orbital distances and stellar ages are presented. 
We conclude that the desert of short-period Neptunes could not be entirely
explained by evaporation of planet atmosphere caused by the radiation
from a host star. For the less massive Neptune-like planet,
the estimated upper limits of the mass loss may be
consistent with the photo-evaporation scenario, while the heavier Neptune-like
planets could not lose the significant mass through this mechanism. We also found
the significant differences between our numerical results and
widely used approximate estimates of the mass loss.
\end{abstract}




\section{Introduction}

The epoch of exoplanet observations and discoveries started more than
twenty years ago. Over this time, more than 3600~exoplanets in about
2700~systems have been identified. The current observational methods
allow not only to detect an exoplanet and to get its orbital characteristics
but also to inspect parameters of its atmosphere \citep{Seager-2010}.
The observations of transits and anti-transits as well as direct observations
of exoplanets made it possible to obtain the spectra of their upper
atmospheres~\citep{Vidal-Madjar-2003,Vidal-Madjar-2004,Linsky-2010,Janson-2010,
Barman-2011}. These spectra, in turn, can be used to determine the chemical
and physical structure of exoplanet atmospheres as well as their dynamics.

Up to now, the most powerful methods of detecting exoplanets are the
transit and radial velocity methods. Both are more effective to identify
exoplanets which are close to their host stars. Therefore, it is natural
that initially discovered exoplanets turned out to have close-in orbits.
Until recently, the most of known planets belongs to ``hot Jupiters'' ---
exoplanets with masses that are comparable to Jupiter's and orbital distances less
than 0.1~AU~\citep{Murray-Clay-2009}. Thanks to the Kepler Space
Telescope, it became possible to detect ``super-Earths'' --- exoplanets
with several Earth masses.

The further analysis of the discovered planet parameters has revealed an
interesting feature. Despite the identification of great number of
Jupiter-like and Earth-like planets at close-in orbits, the number of
``hot  Neptunes'' --- planets with 0.6--18 times of Neptune mass and
orbital periods less than 3 days --- turned out to be very small. The
corresponding region in  the mass-period distribution was assigned as the
``short-period Neptunian desert''~\citep{Szabo-Kiss-2011,Howe-Burrows-2015,Mazeh-2016}. This region can be
hardly explained by selection effects. Indeed, the adopted methods
potentially are more sensitive to more massive planets. While being
successful for identification of super-Earths they should be even more
effective for detecting the short-period Neptunes at the same orbits.
Hence, the desert of short-period Neptunes has likely the physical
origin. The more detailed  analysis of the selection effects and their
relation to the problem of short-period Neptunes can be found
in~\citet{Mazeh-2016}, where the functional expression to describe the
upper and lower boundary of the short-period Neptune desert in the
mass-period diagram was also provided. According to their data, the upper
boundary is represented by the relation $M_p(M_J)=P^{-1.14\pm 0.14}\cdot
10^{0.23 \pm 0.045}$ while the lower boundary can be expressed as 
$M_p(M_J)=P^{0.98}\cdot 10^{-1.85}$, where $P$ is the orbital period
evaluated in days. To explain the origin of the desert the
following hypotheses were mentioned in ~\citet{Mazeh-2016}:
\begin{enumerate}
\item During the protoplanetary evolution the planet migrates 
toward the star from the external parts of protoplanetary disk. This
migration driven by the interaction with the disk stopped near the inner 
edge of the disk where the density is too low to continue pushing the
planet inward. It is suggested that the inner disk radius depends on the
disk mass  being smaller for more massive disks. Thus, the more massive
planets are  presumably formed in more massive disks and settle on
smaller orbital distances.
\item The gaseous planets with Neptune masses would not survive in the vicinity
of host star due to intensive atmosphere outflow caused by stellar radiation.
\end{enumerate}
The second hypothesis is considered in~\citet{Kurokawa-2014, Kaltenegger-2013, Tian-2015}. 
The same process is
observed toward other exoplanets as well. Orbiting close to their parent
stars, the exoplanets with H/He-dominant envelopes are exposed to an
integrated high-energy irradiation that is an appreciable fraction of
their gravitational binding energy~\citep{Lammer-2003, Tian-2005}. Such
high-energy irradiation is typical for the first 100~Myr of the planet's
lifetime~\citep{Jackson-2012}. As a consequence, the H/He-dominant
envelopes can be strongly evaporated by this
irradiation~\citep{Tian-2005, Owen-Jackson-2012, Johnstone-2015}. Such
strong H/He evaporation has been observed from the high-mass planets
HD~209458b~\citep{Vidal-Madjar-2003}, HD189733b~\citep{Lecavelier-2010},
WASP-12b~\citep{Fossati-2010a,Fossati-2010b}, and from the low-mass planet
GJ~436b~\citep{Kulow-2014, Ehrenreich-2015}. Evaporation naturally arises
in planets that are smaller and denser than those at large
separations~\citep{Lopez-2012, Owen-Wu-2013, Lopez-Fortney-2013,
Jin-2014}. For a planet with a low enough mass and a close enough orbit,
its initial low-mass H/He envelope can even be entirely stripped,
leaving behind a naked solid core. The core's mass and density play an
important role in controlling a planet's evolution by setting  the escape
velocity~\citep{Owen-Wu-2013, Lopez-Fortney-2013, Owen-Morton-2016,
Zahnle-Catling-2017}.

The evaporation theory is very useful in the studies of the exoplanet 
demographics. It allows to infer the different
populations of exoplanets. For instance, it predicted the existence of
an ``evaporation valley'', a low-residence region in the radius-period plane
between planets that have been completely stripped and those 
that are able to retain their envelopes with roughly $\approx$1\% in mass.
The evaporation valley was firstly predicted by~\citet{Owen-Wu-2013} using
numerical evolutionary studies for low-mass planets with pure rock (silicate)
cores, and shortly after, by~\citet{Lopez-Fortney-2013} for various core
compositions using a different evaporation model.  This feature can be 
considered also as a radius valley, i.e. a bimodal distribution of planet
radii, with super-Earths and sub-Neptune planets separated by 
a valley at around~$\approx 2 R_{Earth}$. Such a valley was observed recently, 
owing to an improvement in the precision of stellar, and therefore
planetary radii~\citep{Fulton-2017, Van-Eylen-2017}. The evaporation valley
is thus a robust prediction of evaporative driven evolution of close-in 
H/He dominant planets~\citep{Owen-Wu-2013, Lopez-Fortney-2013, Jin-2014, 
Chen-Rogers-2016, Owen-Wu-2017, Lopez-2017}.

The photo-evaporation hypothesis is usually adopted to explain the upper
boundary of the short period Neptunes desert. The theory designed
to explain both upper and bottom boundaries has been presented 
in~\citet{Matsakos-2016}. They argue that the desert may appear in
scenarios of the orbital circularization of planets that arrive at the
stellar vicinity on  high-eccentricity orbits and of their subsequent
tidal angular-momentum exchange with the star.

In our paper, the study of the mass loss of a hot Neptune due to the
hydrodynamical outflow from its atmosphere is presented. This outflow is
induced by the heating of the atmosphere. The main heating  mechanism is
associated with the absorption of stellar XUV (soft X-rays and extreme
ultraviolet) radiation,  in particular, in the range of 
1--100~nm~\citep{Lammer-2003,Lammer-2009,Shaikhislamov-2014,Tian-2005}.
This process was investigated in~\citet{Kurokawa-2014} where the authors
simulated the atmosphere outflow with the 1D semianalytic evolutionary 
model of a sub-Jupiter. To calculate the mass loss due to stellar radiation
they adopted the following formula from~\citet{Lopez-2014}:
\begin{equation}
 \frac{dM_p}{dt}= - \frac{\pi \eta F_{XUV} R^3_{XUV}}{GM_p K_{t}},
 \label{mdoteq}
\end{equation}
where $M_p$ is the planet mass, $F_{XUV}$ the incident XUV stellar flux,
$R_{XUV}$ the radius (evaluated from the planet center) at which the XUV
emission is absorbed, $K_{t}$ the coefficient to account for the tidal
force from the star, and $\eta$ the heating efficiency. Using the
obtained outflow rates and corresponding exoplanet lifetimes,
\citet{Kurokawa-2014} derived the upper boundary of short period
Neptunes desert in period-mass diagram. With the adopted value of heating
efficiency 0.25, the planets below the upper boundary should be evaporated
in less than 2~Gyr. However, in our study~\citet{Shematovich-2014} it was
shown that the mean value of heating efficiency is equal to 0.14 if the
generation of suprathermal electrons is taken into account. Since the
heating efficiency is an important parameter let us describe the heating
of the atmosphere in more detail.

The XUV radiation causes the ionization of atomic hydrogen and helium,
as well as the ionization, dissociation and dissociative ionization
of molecular hydrogen~\citep{Shematovich-2010,Shematovich-2014, Ionov-2015}.
These processes are described by the following equations:
\begin{equation}
\begin{array}{l}
 H_2 + h\nu (e_p) \rightarrow H_2^+ + e_p + (e) \\
 H_2 + h\nu (e_p) \rightarrow H(1s) + H(1s, 2s, 2p) + (e_p) \\
 H_2 + h\nu (e_p) \rightarrow H(1s,2p) + H^+ + e_p + (e) \\
 H + h\nu (e_p) \rightarrow H^+ + e_p + (e) \\
 He + h\nu (e_p) \rightarrow He^+ + e_p + (e),
\end{array}
\label{eq_ioniz}
\end{equation}
where $h\nu$, $e_p$, $e$ are related to XUV-photon, photoelectron
and secondary electron, correspondingly.

The fraction of absorbed photon energy provides ionization
and dissociation while the remaining part is transferred into
kinetic energy of the products, mainly, into kinetic energy
of electrons. If the kinetic energy of the released photoelectron
is high enough, i.e. when it is about ten times higher than
local thermal energy (such electron is called suprathermal),
this electron can stimulate the secondary ionizations and
dissociations. At the same time, the suprathermal photoelectron
can lose its energy during elastic collisions with other particles.
Thus, the part of the photoelectrons energy goes into ionization
and dissociation while the other part provides gas heating.
Correspondingly, the consideration of processes with suprathermal
electrons reduces the amount of energy which goes into heating,
and thus significantly reduces the heating efficiency. 
Moreover, taking into account the suprathermal electrons the
ionization rates will change as well, as shown in~\citet{Ionov-2014}.

The goal of our paper is to check the possibility to explain
the existence of short period Neptunes desert using the model of 
hydrodynamical outflow when appropriate processes with
suprathermal electrons are taken into account. To answer this question
we present the results of numerical simulations of hot Neptune
atmosphere dynamics. We adopt the self-consistent
model by~\citet{Ionov-2017} of hydrogen-helium atmosphere with
suprathermal electrons accounted.

\section{Model}
Our model can be splitted into the kinetic, chemical, and hydrodynamical
parts. Schematically, its structure is shown in~\citet{Ionov-2017}
(Fig. 1, page 388). In the kinetic model, the heating intensity, as well
as the ionization, dissociation, and excitation rates of the atmospheric
species are calculated with the Monte Carlo model. The concentrations of
the atmospheric species are calculated in the chemical model, based on
the reaction rates evaluated in the kinetic model. Using the heating
rates, the variations of atmospheric gas density, velocity,
and temperature are provided by the gas-dynamical model.

The transport and kinetics of photoelectrons in planetary atmosphere were
computed using the Monte Carlo model adopted from~\citet{Shematovich-2008,
Shematovich-2010}. This model includes the reactions~\eqref{eq_ioniz} and
the transport of suprathermal electrons. The energy of a secondary
electron formed during a collision and subsequent ionization is chosen
following the procedure described in~\citet{Marov-1996, Shematovich-2008,
Shematovich-2010}.

With the Monte Carlo method, we compute the rate at which the energy of
stellar XUV radiation and photoelectrons is transformed into the internal
energy during the photoreactions and reactions with secondary electrons.
Separately, we compute the energy of suprathermal photoelectrons that is
transformed into the thermal energy. Using these rates the heating
function of the atmosphere is calculated.

A system of chemical rate equations is solved in the chemical model. The
chemical network includes 19 reactions involving nine components: H,
H$_2$, e$^-$, H$^+$, H$_2^+$, H$_3^+$, He, He$^+$,and  HeH$^+$. The
reaction rates were taken from~\citet{Munoz-2007}.  One of the main
channels for radiative cooling is the emission by H$_3^+$ ion. To
calculate the corresponded cooling function we take the
temperature-dependent emissivity of H$_3^+$ from~\citet{Neale-1996}.

The gas-dynamical model is based on the numerical code described
in~\citet{Pavlyuchenkov-2015}. This 1D code was initially developed to
compute the collapse of a protostellar cloud, and was adapted by us for a
planetary atmosphere. The computations were carried out using Lagrangian
grid, i.e. with moving cell boundaries. In our calculations, we used a
grid that was non-uniform in the mass coordinate: the mass of the cell
falls off with height as a geometric progression with factor 1.03. The
artificial viscosity was introduced into the scheme in order to suppress
oscillations arising in dense layers of the atmosphere. The value of this
viscosity was chosen to be sufficient to suppress non-physical
oscillations.

\section{Model parameters}

Our primary goal is to check the principal possibility for a hot Neptune
to lose the most of its atmosphere due to the heating by stellar
radiation. Therefore, our simulations were intended primarily to
establish the upper limits for atmospheric mass loss rates. We also aim
to compare our calculations with the estimates based on the approximate
formula from~\citet{Kurokawa-2014}.

In our previous paper~\citet{Ionov-2017}, the gravitational potential was
defined by a planet only. However, the outflowing atmosphere is extended
and its dynamics can be affected by the host star as well. Indeed, the
simulations by~\citet{Bisikalo-2013a,Bisikalo-2013b} showed that the
atmospheres of close-in gaseous planets overfill their Roch lobes that
significantly increases the mass loss rate.

So, in this paper we use the approximate spherically-symmetric potential
in the form of the Roche potential in the direction toward the $L_1$
point:
\begin{equation}
\Phi=- \frac{GM_*}{a-r}- \frac{GM_{p}}{r}-0.5 \frac{G (M_*+M_p)}{a^3} \left(r- \frac
{M_{*}}{M_*+M_{p}} a \right)^2,
\label{roche}
\end{equation}
where $a$ is the semi-major axis of the planet, $M_*$ is the stellar
mass, and $M_p$ is the planet mass. This assumption does not recover the
exact 3D nature of the Roche potential but is sufficient to obtain upper
limits of mass loss since the outflow rate is highest toward the $L_1$
point. The same approach was adopted in~\citet{Munoz-2007}. In our 
calculations we assumed that stellar mass is equal to solar mass.

\begin{figure*} 
\begin{center}
\includegraphics[width=110 mm]{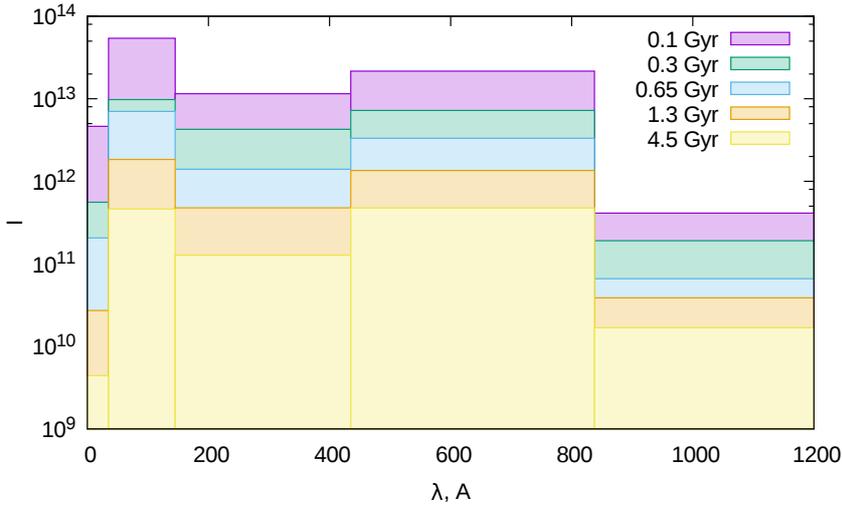}
\caption{The stellar radiation fluxes integrated over
selected wavelength bands. The distributions for different ages
are shown with colors. The wavelength ranges are 1--20, 20--100,
100--360, 360--920 and 920--1180\,\AA.}
\label{spectra}
\end{center}
\end{figure*}

As was noted by~\citet{Ribas-2005}, the intensity of XUV radiation of
young solar-like stars could exceed the present solar intensity in
10--100 times. This should increase the corresponded mass loss rates and
has to be taken into account. Thus, we perform our simulations using not
only the solar spectrum but the spectra of young stars based on data
from~\citet{Ribas-2005}. Namely, we use approximate expressions to
calculate the stellar radiation intensity at 1--20, 20--100, 100--360,
360--920, and 920--1180\,\AA\, for the following stellar ages: 0.1, 0.3,
0.65, and 4.5 (present) billion years. The corresponding fluxes
integrated over selected wavelength bands are shown in Fig.~\ref{spectra}.

\begin{figure*} 
\begin{center}
\includegraphics[width=110 mm]{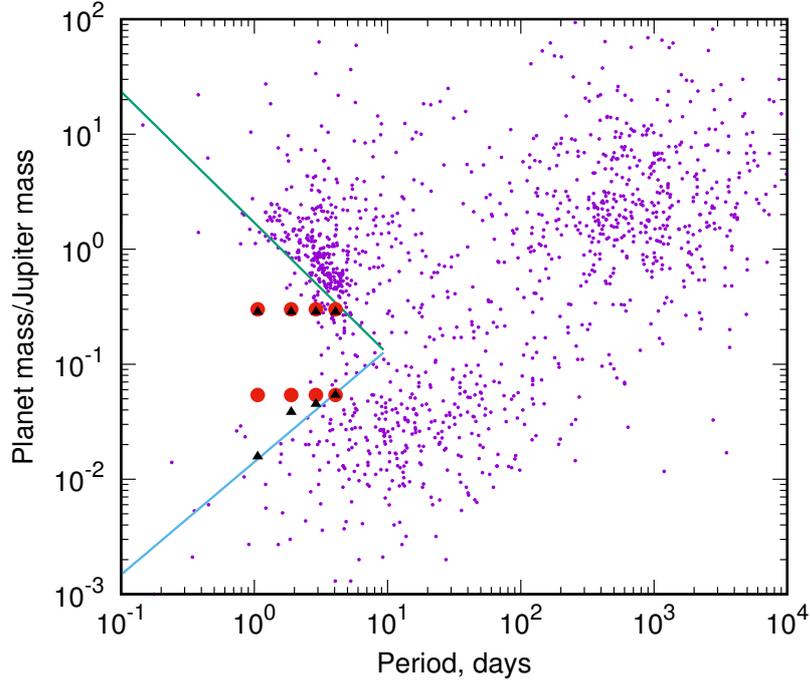}
\caption{The observed exoplanet parameters
extracted from exoplanet database at www.exoplanet.eu
(blue dots), the initial (red circles) and final
(black triangles) positions of the considered models in
the period-mass diagram. The final model positions are
calculated assuming that the planet matter was purely
hydrogen-helium. The boundaries of short-period Neptunian
desert are indicated by lines according to data
from~\citet{Mazeh-2016}.}
\label{hello}
\end{center}
\end{figure*}

Our calculations were performed for two cases of Neptune-like planets.
The first model planet which we name ``heavy'' corresponds to the upper
part of short-period Neptune desert and is six times heavier than
Neptune, i.e. is equal to $6 \times 10^{29}$~g. The second model planet
assigned as ``light'' has the value of Neptune mass and represents the
lower desert boundary. Note, that in the majority of previous studies the
attention was paid presumably to upper desert boundary, see
e.g.~\citet{Kurokawa-2014}. The radii of both planets were calculated
assuming their average densities to be the same as for Neptune. We have
selected 0.05, 0.04, 0.03, and 0.02~AU for planet orbital distances. The
positions of the considered models in the period-mass diagram, together
with the parameters of the observed exoplanets and the estimated
boundaries of Neptunian desert are shown in Fig.~\ref{hello}. Note, that
the models with 0.04 and 0.05~AU correspond to the very edge of the
Neptunian desert.

\section{Mass loss rate}

Let us estimate the mass loss rate from approximate formula~\eqref{mdoteq}.
One of the main parameters in formula~\eqref{mdoteq} is the radius $R_{XUV}$
at which the XUV-radiation is absorbed. The exact value of this parameter
is quite uncertain since the zone of effective XUV absorption is rather
extended. Meanwhile, the mass loss rate in formula~\eqref{mdoteq} is
proportional to the third power of $R_{XUV}$ making the choice of
$R_{XUV}$ important. \citet{Kaltenegger-2013} suggested to
use the radius where the intensity of XUV emission is reduced in $e$
times. Making this choice, however, the intensity at $R_{XUV}$ would
depend on the incident stellar XUV flux. Alternatively, they used the
relation $R_{XUV}=1.4 R_p$, where $R_p$ is the planet radius. In the 
paper by~\citet{Kurokawa-2014} it was suggested to select $R_{XUV}$ as a
radius at which the pressure is equal to $p=10^{-9}$ bar.

\begin{figure*}
\begin{center}
\includegraphics {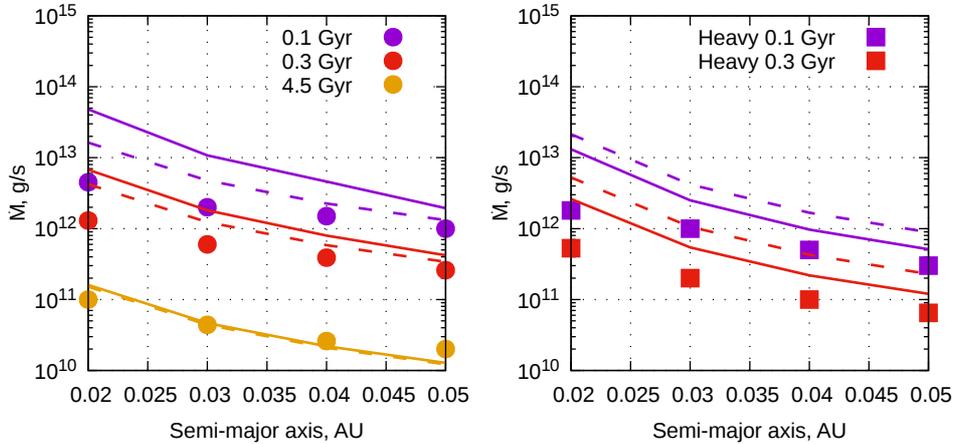}
\caption{The mass loss rate as a function of the planet semi-major axis. The rates
for different star ages are shown with different colors. 
The results for the light and heavy Neptune-like planets
are given in the left and right panels, correspondingly.
The simulated mass loss rates are shown with filled circles and boxes.
The approximate rates calculated with formula~\eqref{mdoteq}
and assuming $R_{XUV}=1.4 R_p$ are shown with dashed lines while 
the solid lines correspond to the case when $\tau(R_{XUV})=1$.}
\label{massloss}
\end{center}
\end{figure*}

The mass loss rates estimated using approximate formula~\eqref{mdoteq} 
and those calculated using our model are shown in Fig.~\ref{massloss}.
The left and right panels correspond to the light and heavy Neptune-like planets,
respectively. The approximate rates calculated assuming $R_{XUV}=1.4 R_p$ 
are shown with dashed lines. The solid lines correspond to the assumption
that $R_{XUV}$ is the radius where the optical depth to XUV radiation is
equal to unity. The optical depth is calculated based on the results of
our simulations. For the case of light Neptune-like planet, the latter
radius exceeds the former one (except for the age of 4.5 billion years
when these radii are nearly the same). Thus, for the light planet, the
approximate mass loss rates for the case $R_{XUV}=1.4 R_p$  are lower
than rates for $\tau(R_{XUV})=1$. For heavy planet the situation is
opposite. The analytic estimates of mass loss rates were calculated with
the heating efficiency $\eta = 0.25$ as in~\citet{Kurokawa-2014}.

From the comparison between approximate curves and points estimated
with our models
one can figure out the following conclusion. For the star with
present solar spectrum, the formula~\eqref{mdoteq} provides a good
fit to the results of numerical simulations. However, for the younger
star the difference between approximate and calculated in our model rates
becomes more significant. For the fixed stellar age, the difference
becomes larger with the decrease of planet semi-major axis. Consequently,
the approximate rates calculated with formula~\eqref{mdoteq} are overestimated
especially during the young stellar ages and closer-in planet orbits,
i.e. in the cases when the mass loss rate is highest. This conclusion
is valid both for the light and heavy cases of the considered
Neptune-like planets.

Let us examine the question whether the calculated mass loss rates can
cause a significant change in the planet mass. To calculate this, we have
to take into account that the mass loss rate from a planet varies in time
following the evolution of the stellar spectrum and luminosity. Given the
limited number of calculated rates, we integrate the total mass loss
$\Delta M$ with the approximate sum:
\begin{equation}
\Delta M = \sum_{i=1}^{5}\dot{M}_i \Delta t_{i}, 
\label{deltam}
\end{equation}
where $\dot{M}_i$ is the mass-loss rate, corresponding to the
{ \it i-}th stellar XUV flux shown in Fig.~\ref{spectra}, 
$\Delta t_i$ is its duration. Note, that this approach is rather
crude and tends to overestimate the total mass loss.

First, let us consider the results obtained for the light Neptune-like
planet model. Despite the fact that outflow rate estimated with our model is lower
than that calculated with formula~\eqref{mdoteq}, it is still sufficient to
cause a significant change of the planet mass. The important question
here is the fraction of the hydrogen-helium envelope in the total mass of
the planet. Under assumption that the hydrogen-helium envelope in the
light Neptune-like planet model has the same mass as in Neptune, all the
light planets placed in the  short-period Neptunian desert will
experience a complete loss of the envelope. If we assume that the
envelope mass is 30\% of the planet mass, the envelope will be completely
evaporated for the orbits with periastris of 0.02 and 0.03~AU. However,
the evaporation of the hydrogen-helium envelope does not mean that the
planet escapes the short-period Neptunian desert. The planet still
should get rid of a substantial fraction of mass (most probably stored in
the water mantle). However, our model does not allow us to follow the
evolution of planet after evaporation of the hydrogen-helium envelope.
Therefore, the question whether the light Neptunes may survive in the 
short-period Neptunian desert requires further consideration.

In case of heavy Neptune-like planet the situation is more certain. The
outflow rates are comparable to that of light planet. This is quite
reasonable since the effect of larger planet surface is compensated by
stronger gravitational acceleration. Being heavier, the planets lose only
5~\% of the initial mass over first 1.5 billion years.

\section{Conclusion}

Based on the obtained results we conclude that the desert of short period
Neptunes could not be reliably explained by the evaporation of planet
atmospheres due to XUV radiation from host star. 

While for the case of
light Neptune-like planet, the calculated upper limits of the mass
outflows are consistent with photo-evaporation scenario,
in the case of heavy Neptune-like planets this mechanism does
not allow to remove the significant fraction of the planet mass.
Some other physical mechanisms like a high-eccentricity migration of planets
that arrive in the vicinity of the Roche limit, where their orbits are tidally
circularized, could be involved for the explanation of the upper boundary
for the short-period Neptunian desert~\citep{Matsakos-2016}.

However, we must note that the calculated outflow rates are, in general,
not fully consistent since we did not follow the evolution of the planet
structure. It is necessary to take into account the change of mass and
radius of the planet during the evaporation. The important effect is that
a young planet may have much larger radius (than that for the evolved
planet). Consequently, the young planet would be affected by stronger
evaporation. These and other evolutionary effects are especially
important in the case of a light Neptune-like planet where the outflow
rates could be significant. We did not take these effects into account
since our current approach to calculate the outflow rates is too
computationally time consuming to be combined with such evolutionary model.

We did not consider a number of processes that could eventually increase
the planet outflow rates. The first one is the flare activity of the 
host star. Indeed, during the star flares the stellar XUV intensity can
be an order of magnitude higher than one in the quite state. As found
by~\citet{Glassgold-2005}, for solar-like star the duration of flashes is
about $10^5$~sec, their X-ray intensity is 10~times higher than in quite
state, and the interval between the flares is about 4--8 days. With the
accounting of the stellar flares, the total XUV energy absorbed by a
planet can be about three times higher than one considered in the quite
conditions. According to formula~\eqref{mdoteq} the outflow rate is
proportional to XUV flux. Thus, the effect of flashes can be responsible
for the ``cleaning'' of the bottom part of short period Neptune desert.
However, this effect can not provide sufficient XUV fluxes to evaporate
heavy Neptune-like planets i.e. to depopulate the top part of the desert.

Another factor that can be important is the coronal activity of the star.
The influence of coronal mass ejections onto stability of a hot Jupiter
atmosphere was studied in~\citet{Cherenkov-2016} and it was shown that
the coronal mass ejection can stimulate the enhanced planet atmosphere
loss. However, it is difficult for us to estimate this effect in case of
hot Neptunes in frames of the adopted spherically symmetric model. One
needs to use 3D hydrodynamical simulations to figure out if coronal mass
ejection could significantly increase the rate of hot Neptune atmosphere
evaporation.

\section{Acknowledgements}
We thank an anonymous referee for her/his suggestions 
that helped us to improve the quality of our paper. 
V. Shematovich acknowledges the support by the Russian
Science Foundation (grant 18-12-00447).


\bibliographystyle{mnras}
\bibliography{stherm} 


\bsp	
\label{lastpage}
\end{document}